\documentclass[prl,showpacs,twocolumn,aps]{revtex4}
\usepackage{graphicx}
\newcommand{\be}{\begin{equation}}
\newcommand{\ee}{\end{equation}}
\newcommand{\bea}{\begin{eqnarray}}
\newcommand{\eea}{\end{eqnarray}}

\newcommand{\etal}{{\it et al.} }
\begin{document}
\title{Low Temperature Thermal Conductivity 
of High Purity YBa$_{2}$Cu$_{3}$O$_{6.99}$ in the Vortex State} 
\author{Wonkee Kim,$^{1}$ F. Marsiglio,$^{1}$ and  J. P. Carbotte$^{2}$}
\affiliation{
$^{1}$Department of Physics, University of Alberta, Edmonton, Alberta,
Canada, T6G~2J1\\
$^{2}$Department of Physics and Astronomy,
McMaster University, Hamilton,
Ontario, Canada, L8S~4M1}
\begin{abstract}
The low temperature thermal conductivity of a $d$-wave superconductor
is considered for arbitrary strength of the impurity potential.
A random distribution of vortices is included within the usual semiclassical 
approach. The vortex-quasiparticle scattering is accounted for 
phenomenologically, in addition to impurity scattering. 
Application of the theory to recent data on high purity 
YBa$_{2}$Cu$_{3}$O$_{6.99}$ shows excellent agreement with experiment,
both in zero field and in the presence of an applied magnetic field.
We find that the strength of the impurity potential required to describe
the data deviates qualitatively from either the unitary or Born limits.
\end{abstract}
\maketitle

Recent measurements of the in-plane thermal conductivity $(\kappa)$
of a high purity
YBa$_{2}$Cu$_{3}$O$_{6.99}$ single crystal \cite{hill}
have revealed disagreement with
the standard $d$-wave theory in several respects. 
The very low temperature $(T)$
variation of $\kappa$ fits neither the unitary nor the Born limit of impurity
scattering theory. In addition, when a magnetic field $(H)$ is applied up
to $13$ Tesla, the $T=0$ data disagree with the predictions of the
usual semiclassical theory of the Doppler shift which accounts for the
supercurrents circulating around the vortex cores. $\kappa$ initially
increases very rapidly reaching almost twice its universal value $\kappa_{00}$
as $H$ increases to a fraction of a Tesla. Beyond this value, 
$\kappa$ remains
almost unchanged up to $13$ Tesla while the semiclassical theory of
K{\"u}bert and Hirschfeld \cite{kh} predicted it should continue to rise as
$\sqrt{H}$. 

Quasiparticles can be created not only thermally  but also 
by the magnetic field. This additional creation of quasiparticles
gives rise to a rapid increase of $\kappa$.
However, as the density of vortices is increased the quasiparticles should
also experience additional vortex-quasiparticle scattering due to
Andreev reflection, and this should have the corresponding effect
of decreasing $\kappa$. It is well-known that vortex scattering has 
important effects on the thermal conductivity of a type II superconductor
\cite{cleary}. However, it was not
included in the work of Ref.\cite{kh}, which applies only to the dirty limit,
where it is assumed that impurity scattering is dominant. To treat the pure
limit where the vortex scattering plays a significant role, 
we follow a semi-phenomenological approach. We use Matthiessen's
rule to add the vortex scattering to the usual impurity term which is left 
unchanged. 

A random distribution of vortices is assumed and a configuration 
average is done as usual. 
After the averaging the vortex scattering is dependent only on the 
single vortex cross section. This quantity has been evaluated by 
Yu \etal \cite{yu}, who gave a simple formula which
is proportional to the magnetic energy $E_{H}$.
Here the proportionality constant will be treated as a phenomenological
parameter. Franz \cite{franz} has also given a dimensional argument
for this form. However, his approach does not give
the correct behavior of $\kappa$ at low temperatures
because in his calculation $\kappa\rightarrow\kappa_{00}$ as $T\rightarrow0$
regardless of $H$, which is inconsistent with experiments.
While Vekhter and Houghton \cite{vekhter} did include
Andreev scattering in their work, they start, however, with an Abrikosov
lattice which introduces extra complications associated with Bloch's theorem.
Their model does not apply to a random distribution and it also
predicts a $\sqrt{H}$ behavior at low temperature. Moreover, their theory
is not well-justified for $H\ll H_{c2}$.
Arguing that the Dirac nodes remain unaffected even in the vortex state
(except for renormalized Fermi and gap velocities), Franz and Vafek \cite{fv}
concluded that for $H_{c1}\ll H\ll H_{c2}$ the thermal conductivity
is independent of $H$. 
This may happen only in a perfect
vortex lattice; however, it is debatable if vortices form a lattice
in a rigorous sense even in a very pure sample.

In the work of Hill \etal\cite{hill} for $H=0$, 
$\kappa/T$ rises out of its universal
limit very fast as a function of temperature, and reaches 
4 times this universal value within
$400mK$. This behavior cannot be understood within either of the two
limits extensively studied in the past \cite{graf}:
Born and unitary limits, corresponding to the impurity potential 
$U_{i}\rightarrow 0$ and $U_{i}\rightarrow \infty$, respectively \cite{note1}.
In this paper we consider explicitly the intermediate case. A dimensionless
parameter $c$ is introduced, as usual, with $c^{2}=1/(U_{i}N(0))$, where
$N(0)$ is the normal state density of states at the Fermi level.
Within a $t$-matrix approximation, the effective scattering in 
the superconducting state can be characterized by $c$ and $\Gamma= n_i/(\pi
N(0))$, where $n_i$ is the impurity density.
We find that the data of Hill \etal \cite{hill} can be fit naturally
with a value of $c\simeq0.014$, which does not belong to either of
the two limiting cases.
The rapid rise in $\kappa/T$ with $T$ also reflects the high purity of
the sample ({\it i.e} small value of $\Gamma$). 

An important feature of 
the impurity scattering is that the scattering rate $\gamma(\omega)$ in the
superconducting state is strongly dependent on frequency and that our
fit to data for YBa$_{2}$Cu$_{3}$O$_{6.99}$ with $T_{c}\simeq 90K$
gives $\gamma(0)=0.29K$,
which is very small in comparison to, for example, the value of the magnetic
energy at $H_{c1}$. This value is of order
$10^{-2}\Delta_{0}$ while $\gamma(0)\simeq 0.0015\Delta_{0}$,
where 
$\Delta_{0}$ is the amplitude of a $d$-wave gap and is 
$2.14T_{c}$ for a $d$-wave superconductor. 
Thus when we include a magnetic field $H$ into the theory
we will need to work in the regime $E_{H}>\gamma(0)$. The limit
$E_{H}<\gamma(0)$ is unrealistic for such pure samples. Nevertheless,
provided $E_{H}/\Delta_{0}$ and $T/\Delta_{0}$ remain small, we can apply
the nodal approximation \cite{durst} 
to reduce complications associated with the calculations.

The superconducting $(2\times2)$ matrix Green's function
in the Nambu notation takes the form
${\hat G}({\bf k},i\omega_{n})=(\tilde{\omega}{\hat\tau}_{0}
+\Delta_{\bf k}{\hat\tau}_{1}+\xi_{\bf k}{\hat\tau}_{3})/
(\tilde{\omega}^{2}-\xi^{2}_{\bf k}
-\Delta^{2}_{\bf k}),$
where ${\hat\tau}$'s are the Pauli matrices in spin space,
$\Delta_{\bf k}$ a $d$-wave order parameter, and $\xi_{\bf k}$ is 
the electronic
energy dispersion in the normal state. The renormalized Matsubara frequencies
$\tilde{\omega}(i\omega_{n})=i\omega_{n}-\Sigma(\tilde{\omega})$ 
include the self-energy
$\Sigma(\tilde{\omega})$, which depends on what interactions are
taken into account. 
Here we consider only the scalar component of
the self-energy assuming that other components can be
neglected or absorbed into $\xi_{\bf k}$ or $\Delta_{\bf k}$\cite{durst}.
In our consideration for a vortex state, it has
two parts: the impurity part $\Sigma_{i}$ and the vortex part
$\Sigma_{v}$. For $\Sigma_{i}$ we use the self-consistent $t$-matrix
approximation while for $\Sigma_{v}$ we take a more phenomenological
approach due to the lack of a detailed microscopic footing.

{\it In the absence of a magnetic field}, $\tilde{\omega}(\omega+i\delta)
=\omega-\Sigma_{i,ret}(\tilde{\omega})$.
Introducing the spectral functions for the diagonal
$\left[G({\bf k},i{\omega})\right]$ and off-diagonal 
$\left[F({\bf k},i{\omega})\right]$ components of 
${\hat G}({\bf k},i{\omega})$,
$A({\bf k},\omega)=-2\mbox{Im}\left[G({\bf k},\omega+i\delta)\right]$ and
$B({\bf k},\omega)=-2\mbox{Im}\left[F({\bf k},\omega+i\delta)\right]$,
respectively, we obtain the thermal conductivity at $T$
\bea
\frac{\kappa (T)}{T}&=&\frac{1}{T^{2}}\sum_{\bf k}(v^{2}_{f}+v^{2}_{g})
\nonumber\\
&\times&
\int\frac{d\omega}{2\pi}\left[A^{2}({\bf k},\omega)-B^{2}({\bf k},\omega)
\right]\omega^{2}\left(-\frac{\partial f}{\partial\omega}\right)\;,
\eea
where $v_{f(g)}$ is the Fermi (gap) velocity and $f(\omega)$ is the
Fermi function. As we mentioned earlier, application of the nodal approximation
allows us to carry out the summation over ${\bf k}$ and yields
\be
\frac{\kappa (T)}{T}=\frac{1}{\pi^{2}}\frac{1}{T^{2}}
\left(\frac{v_{f}}{v_{g}}+\frac{v_{g}}{v_{f}}\right)
\int d\omega{\cal A}(\omega)
\omega^{2}\left(-\frac{\partial f}{\partial\omega}\right)\;,
\label{thermal_cond}
\ee 
where 
${\cal A}(\omega)=1+\left[h(\omega)+\frac{1}{h(\omega)}\right]
\arctan[h(\omega)].$ Here
$h(\omega)$ is defined as
$h(\omega)=(\omega-\mbox{Re}\left[\Sigma_{i,ret}\right])/
\gamma(\omega)$ with $\gamma(\omega)=-\mbox{Im}\left[\Sigma_{i,ret}\right]$.
In order to calculate the thermal conductivity in Eq.~(\ref{thermal_cond}), 
we need to determine the self-energy $\Sigma_{i,ret}$
self-consistently. In the $t$-matrix approximation $\Sigma_{i,ret}$ can be written as
$\Sigma_{i,ret}({\tilde\omega})=
\Gamma G_{0}({\tilde\omega})/\left[c^{2}-G^{2}_{0}({\tilde\omega})\right]$,
where $G_{0}
=[2\pi N(0)]^{-1}\sum_{\bf k}\mbox{Tr}[{\hat\tau}_{0}{\hat G}_{ret}]$. 
As we described, $c$ here
controls the strength of the impurity scattering. 
Applying the nodal approximation, we obtain\cite{ewald}
\be
G_{0}(\tilde{\omega})\simeq
\frac{2}{\pi}\left[
\frac{\tilde{\omega}}{\Delta_{0}}
\ln\frac{\tilde{\omega}}{4\Delta_{0}}-
i\frac{\pi}{2}\frac{\tilde{\omega}}{\Delta_{0}}\right]\;.
\label{G_0}
\ee
Substituting Eq.~(\ref{G_0}) into $\Sigma_{i,ret}$, we compute 
$\mbox{Re}[\Sigma_{i,ret}]$ and
$\mbox{Im}[\Sigma_{i,ret}]$ self-consistently. In the presence of a magnetic
field, the results for $\Sigma_{i,ret}$ remain unchanged. 

Since ${\cal A}(\omega)\rightarrow2$ as $T\rightarrow0$, it is easy to show that
$\kappa(T)/T$ approaches its universal value, 
$(2/3)\left[v_{f}/v_{g}+v_{g}/v_{f}\right]\equiv\kappa_{00}/T$.
In Fig.~1 we show results for $\kappa(T)/T$ normalized by the universal value
for different values of $c$.
The figure is motivated by the data of Hill \etal on a very
pure sample of YBa$_{2}$Cu$_{3}$O$_{6.99}$. Their experimental fit
appears as solid circles corresponding to $(0.16+3.0T^{2})/0.16$
\cite{hill}.  We should note that this data requires
subtraction of a phonon contribution; this can be done in a variety of
sensible ways, which leads to an uncertainty in the electronic part of
several percent \cite{taillefer}. The solid curve shown, a theoretical calculation
with $c=0.014$, is thus in excellent agreement with the data.
While there are two
parameters in the impurity theory, $\Gamma$ and $c$, a constraint was
applied to all fits: namely, $\gamma(0)$ is set to be $0.29K$. This 
constraint comes from consideration of finite as well as zero magnetic
field data as will be explained later. As seen in Fig.~1 
the unitary limit does not fit the data at finite $T$. 
On the other hand
we can achieve reasonable agreement with the data for $c=0.014$ and
$\Gamma/\Delta_{0}\simeq5.05\times10^{-5}$ over the entire temperature
range shown in Fig.~1.
This corresponds to a very clean sample. 
If $\Gamma$ were to be interpreted as the normal state impurity scattering
rate, the corresponding mean free path would be anomalously large.
We take this as an indication that for this sample 
the underlying normal state, 
on which the superconductivity is built, may not be an ordinary
Fermi liquid.
\begin{figure}[tp]
\begin{center}
\includegraphics[height=2.6in]{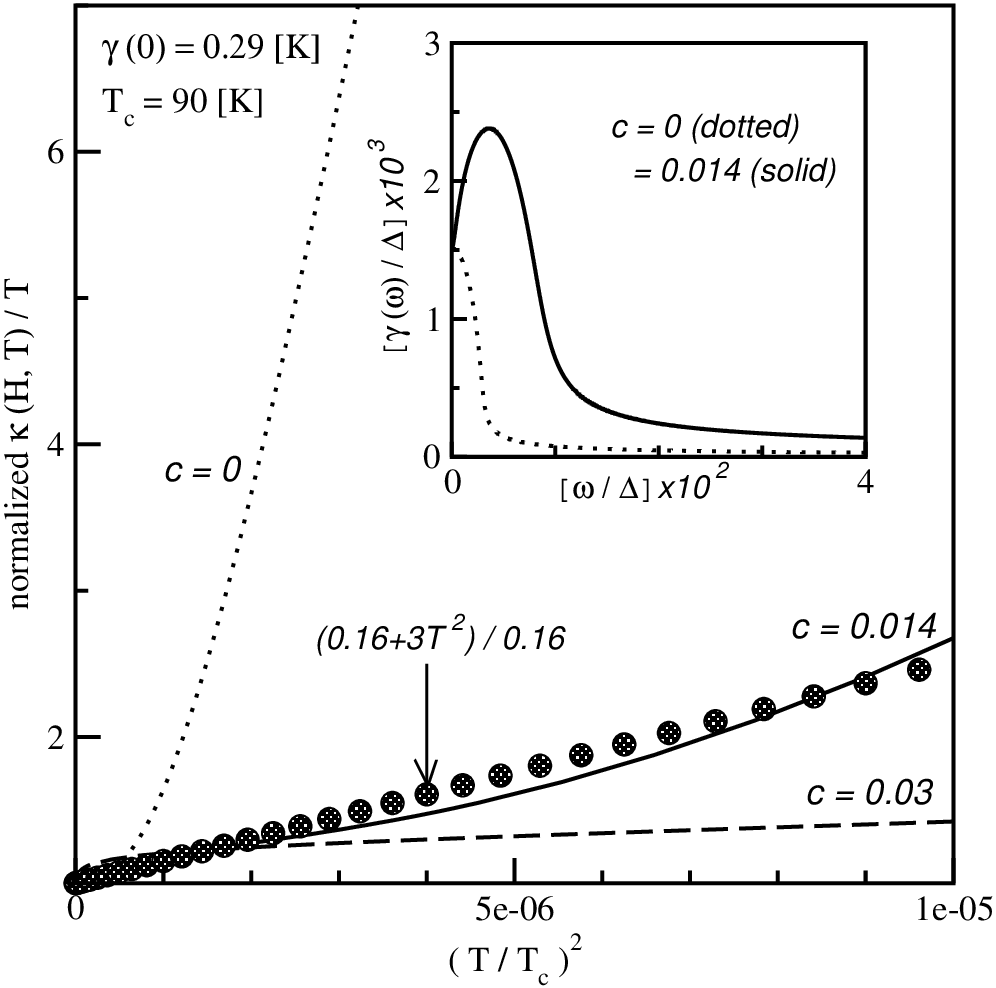}
\caption{The thermal conductivity $\kappa(T)/T$ in units of its universal value
as a function of $(T/T_{c})^{2}$ in the very low $T\ll T_{c}$ range.
In all cases, $T_{c}=90K$ and $\gamma(0)=0.29K$. The impurity
concentration parameter $\Gamma$ varies as $c$ is changed.
Each curve is labeled by the $c$ value used.
The experimental fitting appears as solid circles.
The inset shows the self-consistent results for $\gamma(\omega)$ for
$c=0$ (dotted) and $c=0.014$ (solid curve).
}
\end{center}
\end{figure}

We note that the theory does not give an exact straight line while the
experimental data has been parameterized with a linear fit \cite{hill}.
However, it should be remembered that to obtain the electronic
thermal conductivity from experiment it was necessary to subtract a 
considerable phonon contribution. Uncertainty remains about this subtraction
as we will elaborate on later and so the small disagreement that remains
between theory and experiment is not a serious limitation on the fit
obtained. 
In the inset of Fig.~1 we show $\gamma(\omega)$ obtained for $c=0.014$
(solid curve) compared with the unitary limit (dotted curve). The curves are
completely different. They both start with $\gamma(0)/\Delta_{0}=0.0015$
by design but for finite $\omega$ the solid curve rises while the 
dotted curve drops monotonically. These differences lead to the large 
difference in the $T$ dependence of $\kappa(T)/T$ shown in Fig.~1.

{\it In the presence of a magnetic field}.
Next we wish to include the effect on the 
thermal conductivity of an external magnetic field $H$ applied
along the $c$-axis.
To this end we need to know the self-energy of a quasiparticle 
due to vortices $\Sigma_{v}$ as well as for the impurities. Formally speaking,
the semiclassical approximation takes into account an approximate
real part of the self-energy: $\mbox{Re}\left[\Sigma_{v}\right]\simeq
{\bf v}_{s}({\bf r})\cdot{\bf k}$
where ${\bf v}_{s}({\bf r})$ is the superfluid velocity at a position ${\bf r}$
measured from the center of a vortex core. Here ${\bf r}$ is restricted
to the vortex unit cell and an average over ${\bf r}$ is to be carried out.
For the imaginary part of $\Sigma_{v}$, we consider the Andreev reflection as
a possible mechanism. When a quasiparticle moves from ${\bf r}$ to ${\bf r}'$,
the Andreev reflection takes place effectively 
if $\sqrt{\xi^{2}_{\bf k}+\Delta^{2}_{\bf k}}+{\bf v}_{s}({\bf r})\cdot{\bf k}
=|\Delta_{\bf k}|+{\bf v}_{s}({\bf r}')\cdot{\bf k}$. It has been 
shown \cite{yu}
that for nodal quasiparticles the vortex scattering rate $\gamma_{v}$
is determined mainly
by the intervortex distance $R$; 
namely, $\gamma_{v}\approx v_{f}/R$. 
This implies that within the nodal approximation
a vortex plays the role of a defect or a scattering center.
Since the intervortex distance $R\approx\sqrt{\Phi_{0}/H}$, where $\Phi_{0}$
is a flux quantum, and 
the magnetic energy $E_{H}$ is given by $E_{H}=a\sqrt{\pi/2}v_{f}
\sqrt{H/\Phi_{0}}$, where $a$ is a parameter of order unity\cite{kh},
we can write $\gamma_{v}=bE_{H}$, 
which defines a vortex scattering parameter $b$.
Instead of evaluating the microscopic quantities which determine the size of $b$
from first principles, we propose to treat it as a phenomenological parameter which
we can use to fit the zero temperature data of the thermal conductivity
as a function of $H$.
\begin{figure}[tp]
\begin{center}
\includegraphics[height=2.6in]{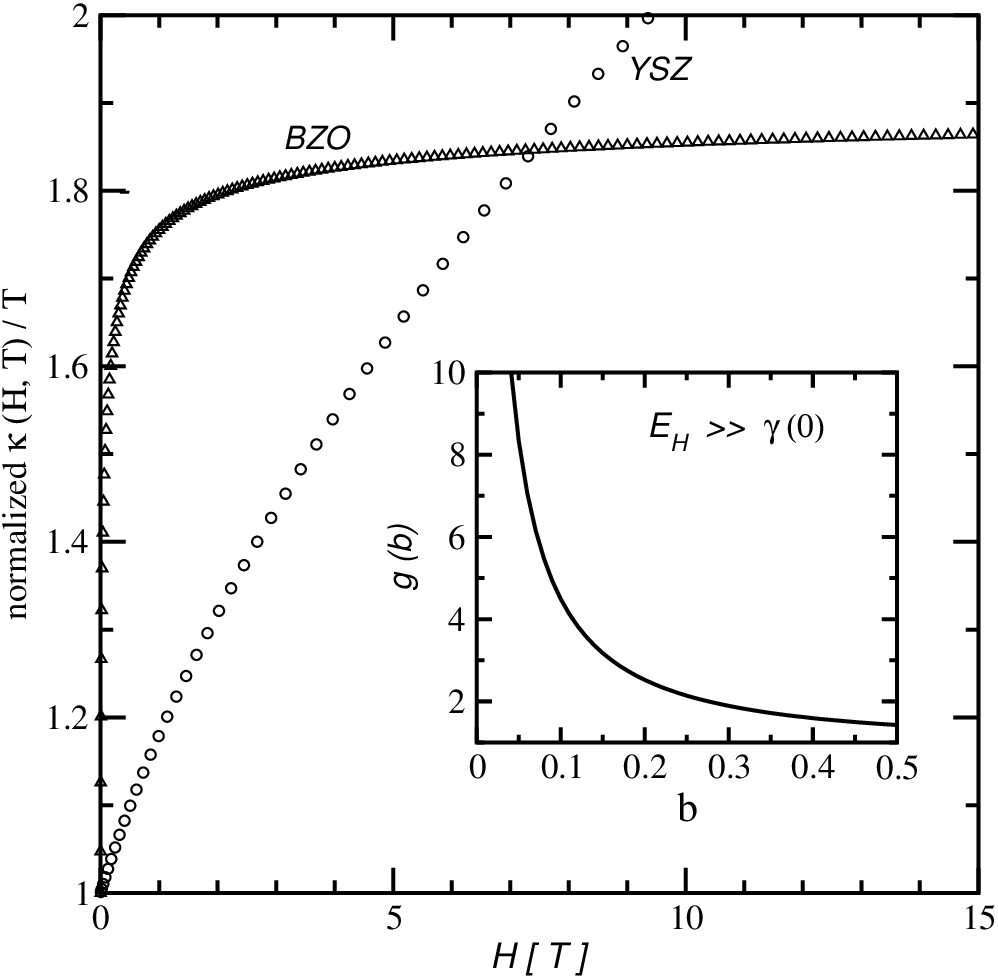}
\caption{
The normalized $\kappa(T)/T$ as a function of $H$ for the BZO sample (triangle)
and the YSZ sample (circle). 
The inset gives the variation of
the magnetic field saturation function $g(b)$ vs. $b$. (See text for more
explanation).
}
\end{center}
\end{figure}

An averaged value of the thermal conductivity over the vortex distribution
can be evaluated as usual: 
$\kappa(H,T)=\int{}d\epsilon{\cal P}(\epsilon)~\kappa(\epsilon,T)$,
where ${\cal P}(\epsilon)=A^{-1}
\int{}d{\bf r}~\delta\left[\epsilon-{\bf v}_{s}({\bf r})\cdot{\bf k}\right]$
is the vortex distribution within the area of a vortex unit cell $A$. 
The appropriate average (series or parallel) depends on the direction of
the field as discussed by K{\"u}bert and Hirschfeld\cite{kh}. 
We will not deal directly
with these issues here. We made numerical calculations assuming
series and parallel arrangements for the thermal conductivity but have
found no qualitative differences. Now the thermal conductivity in a vortex
state becomes
\bea
\frac{\kappa (H,T)}{T}&=&\frac{1}{\pi^{2}}\frac{1}{T^{2}}
\left(\frac{v_{f}}{v_{g}}+\frac{v_{g}}{v_{f}}\right)
\int d\epsilon{\cal P}(\epsilon)
\nonumber\\
&\times&
\int d\omega{\cal A}(\omega,\epsilon)
\omega^{2}\left(-\frac{\partial f}{\partial\omega}\right)\;,
\label{thermal_cond_H}
\eea
where
${\cal A}(\omega,\epsilon)=1+\left[h(\omega,\epsilon)+\frac{1}{h(\omega,\epsilon)}\right]
\arctan[h(\omega,\epsilon)].$
Note the similarity between Eqs.~(\ref{thermal_cond}) and (\ref{thermal_cond_H}).
The effects of the Doppler shift and the Andreev scattering appears in
$h(\omega,\epsilon)$; 
$h(\omega,\epsilon)=\left(\omega-\mbox{Re}\left[\Sigma_{i,ret}\right]-\epsilon\right)
/\gamma_{t}(\omega)$,
where $\gamma_{t}(\omega)=
\gamma(\omega)+bE_{H}$.

Taking the limit of $T=0$ in Eq.~(\ref{thermal_cond_H}) leads to a simplified expression 
as follows:
$\kappa(H,0)=(\kappa_{00}/2)\int{}d
\epsilon {\cal P}(\epsilon){\cal A}(0,\epsilon)$ with
${\cal A}(0,\epsilon)=
1+\left[{\epsilon}/{\gamma_{t}(0)}
+{\gamma_{t}(0)}/{\epsilon}\right]
\arctan\left[{\epsilon}/
{\gamma_{t}(0)}\right]$.
To proceed further some vortex distribution function ${\cal P}(\epsilon)$
needs to be specified. 
We consider the Gaussian distribution;
${\cal P}(\epsilon)=\exp\left[-\epsilon^{2}/E^{2}_{H}\right]/
(\sqrt{\pi}E_{H})$,
which is believed to be favorable for Andreev scattering \cite{distribution}.
Now we obtain the thermal conductivity as $T\rightarrow0$:
\bea
\kappa(0,H)&=&\frac{\kappa_{00}}{2}\int{}dx\frac{e^{-x^{2}}}{\sqrt{\pi}}
\nonumber\\
&\times&
\left\{1
+\left[h(0,x)+\frac{1}{h(0,x)}\right]
\arctan\left[h(0,x)\right]\right\}\;
\eea
where $h(0,x)=x/\left[\gamma(0)/E_{H}+b\right]$. For $E_{H}\gg\gamma(0)$,
$h(0,x)\simeq x/b$ and, thus, $\kappa(0,H)/\kappa_{00}\simeq g(b)$,
where $g(b)=(1/2\sqrt{\pi})\int{}dxe^{-x^{2}}\left[1+(x/b+b/x)\arctan(x/b)\right]$.
Consequently, the magnetic field saturated value of $\kappa(0,H)$ is $g(b)\kappa_{00}$
and not $\kappa_{00}$ as found in Ref.\cite{franz}.

The function $g(b)$ is shown in the inset of Fig.~2. Note that as $b\rightarrow0$,
$g(b)$ goes to infinity as $\sqrt{\pi}/(4b)$ and no saturation occurs in this case.
This also corresponds to the dirty limit case considered in Ref.\cite{kh}
where it is assumed that $\gamma(0)$ is much bigger than $bE_{H}$. 
In the main frame of Fig.~2 we show our results for the normalized
$\kappa(H,T)/T$ vs $H$ at $T=0$ for the two samples considered by Hill \etal,
a very pure sample denoted by BZO with 
$T_{c}\simeq90K$ (triangle) and an impure sample denoted by YSZ
(circle).
In both cases the agreement with the data 
is excellent (compare with Fig.~3 of Ref. \cite{hill}).

For the BZO sample with $\gamma(0)\simeq0.29K$ the value of $b$ obtained 
in the fit is $b=0.3$ assuming $v_{f}\simeq 10^{7}cm/s$
and $a=1/2$ so that 
$E_{H}/\Delta_{0}\simeq0.036\sqrt{H}~$T$^{-1/2}$\cite{chiao}.
Using $H_{c1}\approx 30$mT of YBa$_{2}$Cu$_{3}$O$_{x}$ with $T_{c}\simeq91K$
\cite{handbook}, we have $E_{H}/\Delta_{0}\simeq0.006\Delta_{0}$, which is
greater than $\gamma(0)$.
For the dirty
sample (YSZ) with $T_{c}\simeq93K$, the estimated value of
$\gamma(0)$ is about $13~K$ \cite{hill}.
Using these parameters with $v_{f}\simeq2\times10^{7}cm/s$,
we found that no vortex
scattering parameter needs to be introduced to fit the data; namely,
$b=0$. In this case the magnetic energy at a few Tesla is still of the
order of $\gamma(0)$ and so the data is not sensitive to a small value of $b$.

In Fig.~3 we show results for the normalized $\kappa(H,T)/T$
in units of the universal value as a function of $(T/T_{c})^{2}$.
The solid curve reproduces our $H=0$ result for $c=0.014$. The different
symbols represent different values of $H$; namely, $H=1$T (open circle),
$4$T (open square), and $9$T (open diamond). Note that within $10\%$ or 
less the saturated value of $\kappa(H,T)/T$ due to $H$ is temperature 
independent. This justifies the process used in Hill \etal to subtract out the
phonon contribution from their raw data which includes both an electronic and
a phonon part. We also emphasize that
the crossing with the $H=0$ curve occurs around 
$(T/T_{c})^{2}=5\times10^{-5}$ or $T\simeq0.2K$, which agrees well with
experiment.
We point out that 
to get this agreement we insisted on $\gamma(0)\simeq0.29K$ and
the crossing temperature $T^{*}\simeq0.2K$ is close
to $\gamma(0)$ so that it is the crossing that determines 
this parameter in our approach. 
For $b=0$ the crossing cannot occur.
This implies that both the crossing itself and the saturated 
value of $\kappa(0,T)$ show the significance of the 
vortex scattering.
In the inset we show a few more results
for small values of $H(\gtrsim H_{c1})$. This inset shows that for a weak
field the thermal conductivity retains significant temperature dependence
although considerably reduced from the $H=0$ case. As one can see in this
inset or the main frame of Fig.~2, at $T=0$ the thermal conductivity
depends strongly on $H$ for $H_{c1}\lesssim H <1$T. This can be seen
in the experiments. 
\begin{figure}[tp]
\begin{center}
\includegraphics[height=2.6in]{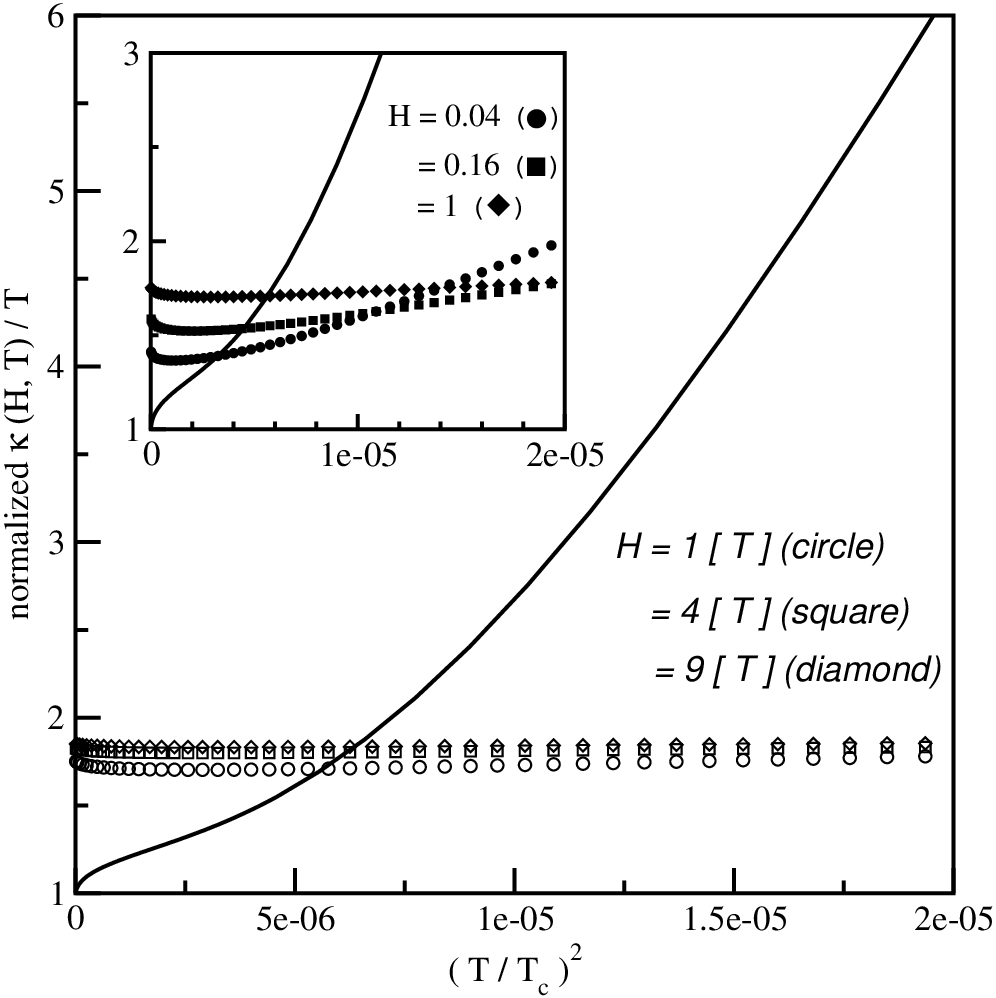}
\caption{
The normalized $\kappa(T)/T$ as a function of $(T/T_{c})^{2}$.
The solid curve is our result for $H=0$ with $c=0.014$ reproduced
from Fig.~1. The open circle, square, and diamond are for
$H=1$, $4$, and $9$T, respectively. The inset shows additional
results for $H=0.04$T, $0.16$T as well as $H=0$ and $H=1$T
for comparison.
}
\end{center}
\end{figure}

In summary
we have given a theory of the low temperature thermal conductivity of a 
$d$-wave superconductor using a self-consistent treatment of 
impurity scattering
intermediate between the unitary and Born limits. We have also 
included the effects of a magnetic field in the range
$H_{c1}<H\ll H_{c2}$ within the semiclassical approximation which
takes into account the Doppler shift due to the circulating
supercurrent around the vortex cores. Assuming a random distribution of 
vortices, the Andreev scattering from the vortices is considered and this
leads to saturation in the thermal conductivity at $T=0$ for a sufficiently
large value of $H$. We find good qualitative and quantitative agreement in all
respects with recent data on a very pure sample of 
YBa$_{2}$Cu$_{3}$O$_{6.99}$. 
We also verify that the phonon contribution
can be subtracted out using  
the thermal conductivity for a sufficiently high magnetic field.

\begin{acknowledgments}
We wish to thank R. Hill and L. Taillefer for helpful correspondence
and for sending us a preliminary draft of Ref.\cite{hill} before publication.
This work was supported in part by the Natural Sciences and Engineering
Research Council of Canada (NSERC), by ICORE (Alberta), and by the 
Canadian Institute for Advanced Research (CIAR).
\end{acknowledgments}

\bibliographystyle{prl}

\begin{thebibliography}{1}
\bibitem{note1} Assuming $N(0)$ is inversely proportional to the
electronic bandwidth $W$, one can write $c^2 \approx W/U_i$;
therefore the unitary limit ($c = 0$) corresponds to the extreme
situation $U_i >> W$.
\end{thebibliography}

\end{document}